\documentclass[superscriptaddress,hyperref,twocolumn]{revtex4-2}
\usepackage{graphicx}
\usepackage{amsmath,amsthm,amssymb,dsfont}
\usepackage{mdframed}

\begin{document}

\title{Bell's theorem allows local theories of quantum mechanics}
\author{Jonte R. Hance}
\email{jonte.hance@bristol.ac.uk}
\affiliation{Quantum Engineering Technology Laboratories, Department of Electrical and Electronic Engineering, University of Bristol, Woodland Road, Bristol, BS8 1US, UK}
\author{Sabine Hossenfelder}
\affiliation{Frankfurt Institute for Advanced Studies, Ruth-Moufang-Str. 1, D-60438 Frankfurt am Main, Germany}

\begin{abstract}
A recent Nature Physics editorial (\textit{Nat. Phys.} (2022) \textbf{18}, 961) falsely claims ``any theory that uses hidden variables still requires non-local physics.'' We correct this claim and explain why it is important to get this right.
\end{abstract}

\maketitle

A recent Editorial in Nature Physics makes the valuable point that with so much of modern science and technology dependent on quantum mechanics, trying to understand its foundations is time well spent \cite{NatPhys2022Survey}. However, we see the need to correct an unfortunate mistake in the Editorial, which is the statement:
\begin{quote}\textit{``One route out of this paradox could be an undetected, so-called hidden variable associated with both particles that underlies the correlated behaviour. However, John Stewart Bell proved that such an approach cannot explain the quantum mechanical outcomes. Any theory that uses hidden variables still requires non-local physics.''}
\end{quote}
This passage reflects a common misconception \cite{Schlosshauer2013Attitudes,Sujeevan2016Survey} that Bell's theorem proves that hidden variable theories that reproduce quantum mechanics require nonlocal physics. In fact, locally causal completions of quantum mechanics are possible, if they violate the assumption that the hidden variables do not in any way depend on measurement settings.

There is no independent evidence that this assumption---commonly known as statistical independence---is fulfilled for quantum systems. As a consequence, the observed violations of Bell's inequality can be said to show that maintaining local causality requires violating statistical independence. We wish to stress that this is not merely an issue of interpretation. The statistical independence assumption is mathematically necessary for the formulation of Bell-type inequalities \cite{Redhead1987Incompleteness,Vervoort2013Bell,chen2021bell}.

Types of hidden variables theories which violate statistical independence include those which are superdeterministic \cite{Hossenfelder2020Rethinking}, retrocausal \cite{Wharton2019Reformulations}, and supermeasured \cite{Hance2022Supermeasured}. Some have dismissed them on metaphysical grounds, by associating a violation of statistical independence with the existence of ``free will'' or ``free choice'' and then arguing that these are not assumptions we should give up. 

It is, in hindsight, difficult to understand how this association came about. We believe it originated in the idea that a correlation between the hidden variables and the measurement setting would somehow prevent the experimentalist from choosing the setting to their liking. However, this is mistaking a correlation with a causation. And any serious philosophical discussion of free will acknowledges that human agency is of course constrained by the laws of nature anyway. For this reason, the mathematical assumption of statistical independence bears no relevance to the philosophical discussion of free will.

In our view, this oversight and the unwillingness to consider theories without statistical independence may be the reason we do not yet have a locally causal theory for the foundations of physics that is consistent with general relativity. Understanding the implications is even more important now that the experimentally observed violations of Bell’s inequality have been awarded the 2022 Nobel Prize in Physics. Contrary to what is often stated, these observations do not demonstrate that ``spooky action at a distance'' is real and nature therefore non-local. Rather, the observations show that if nature is local, then statistical independence must be violated. We should therefore look for independent experimental evidence that can distinguish the two different options: non-locality and statistical independence, or locality and violations of statistical independence.

\textit{Acknowledgements -} JRH acknowledges support by the University of York's EPSRC DTP grant EP/R513386/1, and the Quantum Communications Hub funded by EPSRC grants EP/M013472/1 and EP/T001011/1. SH acknowledges support by the Deutsche Forschungsgemeinschaft (DFG, German Research Foundation) under grant number HO 2601/8-1.

\bibliographystyle{unsrturl}
\bibliography{ref.bib}
\end{document}